# Time to timestamp: opportunistic cooperative localization from reception time measurements

Fabio Ricciato
Austrian Institute of Technology

**Abstract**

We present a general framework for improving and extending GNSS-based positioning by leveraging opportunistic measurements from legacy terrestrial radio signals. The proposed approach requires only that participating nodes collect and share *reception timestamps* of incoming packets and/or other reference signals transmitted by other fixed or mobile nodes, with no need of inter-node synchronization. The envisioned scheme couples the idea of cooperative GNSS augmentation with recent pioneering work in the field of time-based localization in asynchronous networks. In this contribution we present the fundamental principles of the proposed approach and discuss the system-level aspects that make it particularly appealing and timely for Cooperative ITS applications, with the goal of motivating further research and experimentation in this direction.

## 1. Introduction & motivations

In this paper we present a framework approach to integrate Global Navigation Satellite Systems (GNSS) and legacy terrestrial wireless networks (short-range and/or long-range) in order to (1) improve the localization accuracy for nodes equipped with on-board GNSS receiver and (2) extend the localization function to wireless nodes outside GNSS coverage and/or without onboard GNSS module. The proposed approach is "opportunistic" since it may exploit legacy wireless communication signals from fixed infrastructure and/or between cooperating mobile nodes, and is particularly appealing for applications in the field of Cooperative Intelligent Transportation Systems (C-ITS). Our approach builds upon recent work in the field of time-based localization in asynchronous network [Nag11,Col14,Fac14], whose principles are briefly presented hereafter, combined with the concept of cooperative positioning [Gar12,Ami14]. The requirements for practical implementation are minimal: we demand only that the participating devices are capable of recording reception timestamps for legacy communication signals, and that have the means to share these data along with their initial position (if available) as provided by GNSS.

In this work we do not provide a single ready-made solution for a specific scenario or use-case, but rather describe a more general reference framework and the underlying principles of the proposed solution space. Our goal is twofold. On the side of the research community, we indicate a coherent research agenda aimed at attracting attention onto the (mostly unexplored to date) field of opportunistic and cooperative *time-based* localization between asynchronous nodes, highlighting some prominent directions for investigation and experimentation. On the side of the industry, we aim at drawing attention on the potential value (in terms of costs *vs.* benefits) of supporting more accurate timing measurement in commercial-off-the-shelf (COTS) receivers for terrestrial communication, and making these measurements available via standard interfaces to system integrators. In other words, we envision the future "commoditization" of precise receiver timestamping function in COTS devices, similarly to what has happened for the Received Signal Strength Indicator (RSSI), in order to unlock the potential of opportunistic time-based multi-radio localization.

The paper is organized as follows. We start in Sec. 2 by representing some fundamental trends in wireless systems technology that collectively motivate the appeal, timeliness and practical viability of the proposed method for real-world adoption. In Sec. 3 we describe in a tutorial manner the fundamental principles of time-based localization in asynchronous networks. Along the way, we review relevant pioneering work and highlight a number of relevant research directions for the scientific community. Finally, in Sec. 4 we conclude with some simple recommendations for industry players.

## 2. Current trends

### Trend 1: Radios coming together

Generally speaking, radio signals can be used to perform different functions: communication, localization and navigation, sensing and imaging, to name the most popular nowadays. Historically, the vast majority of radio technologies are designed to serve only one function, e.g. communication *or* localization. Dedicated single-function radio systems are easier to design and allow optimizing each component (signal format, protocol, device etc.) for the function of interest. However, real-world applications and use-cases often involve a blend of multiple functions, for example communication *and* localization. The current dominant paradigm is to use multiple single-function systems: for example in the field of C-ITS one common option is to rely on a GNSS module for localization, to DSRC/IEEE 802.11p module for local-area communication, and to a UMTS/LTE module for wide-area communication. In this way, multiple radio modules developed independently for different functions end up together, co-located in the same On-Board Unit (OBU). Similar considerations can be done for smartphones. Given this scenario, it is quite natural to ask whether some form of tighter integration or inter-working between these modules could be exploited to improve the overall system performance. As we will show hereafter, the answer is positive (at least) as far as the positioning accuracy is concerned, meaning that there is a (mostly unexplored) potential to improve the positioning accuracy through clever interworking of these different systems.

### Trend 2: Reusing Radios

It is possible that in the mid-term future we will see a proliferation of wireless technologies designed natively to support multiple functions. One example in this direction is given by the Ultra Wide Band (UWB) technology wherein the signal format, protocol and device architecture have been designed and standardized from the beginning to serve both communication and ranging functions. In principle, in C-ITS applications one could develop systems that use a single waveform to let the vehicle sense and communicate with other vehicles. However, along the path from (current) single-function systems towards (future) multi-function systems there are opportunities to reuse "opportunistically" systems that were developed for some "native" function X in order to perform also a different "supplementary" function Y. For example, certain opportunistic radars (also called "green radars") exploit legacy signals from existing broadcasting towers. Of particular interest for practical applications are those cases where the supplementary function Y can be performed with small or null adaptation of the legacy devices, therefore at (almost) zero additional cost. The most popular example is probably represented by radio localization methods based on the Received Signal Strength Indicator (RSSI) in WiFi, Bluethoot, IEEE 802.15.4 and other wireless systems: strictly speaking, RSSI is not required to perform the native function of these systems, i.e., communication. Nevertheless, RSSI is now a "commodity" function supported by most (if not all) commercial devices, and that has enabled the implementation of *coarse* localization capabilities in these systems. Along a similar line of reasoning, we claim here that timing measurements on the receiver side, i.e., accurate timestamping, is a small but powerful ingredient holding the potential to introduce more accurate localization capabilities especially in multi-radio nodes (for instance smartphones and OBUs) without requiring clock synchronization between them.

### Trend 3: Cooperation

Before turning of the century the dominant paradigm for mobile radio systems was asymmetric and vertical: it would involve mobile end-terminals "served" by fixed infrastructure nodes with different capabilities. This model reflects the asymmetry of the traditional *service-provisioning model*, wherein provider and consumer are distinct entities. In the last decade however, the peer-to-peer paradigm has made his way, first in computer applications and then in the real of wireless communications. Systems based on the horizontal cooperation between peer nodes that *collectively build the service* have been developed and accepted. In many scenarios, and prominently in the C-ITS field, hybrid radio systems are envisaged to include a combination of horizontal and vertical components, i.e., vehicle-to-vehicle (V2V) and vehicle-to-infrastructure (V2I). The application of the peer-to-peer concept for GNSS augmentation has been proposed earlier in [Gar12] (see also [Ami14] and references therein) where it was shown that *the fusion of initial GNSS positions and ranging measurements* between the nodes bears the potential to *improve* the position estimates (for GNSS-enabled nodes) and at the same time *extend* the localization function to non-GNSS-enabled nodes. However, the system model considered in those papers assumes that *nodes are equipped with additional ranging capabilities*. The latter can be delivered by external dedicated sensors, e.g. radar, or by implementing dedicated ranging procedures on the wireless communication channel, e.g. two-way Time-of-Arrival. Instead, we are interested in methods that reuse the existing communication signals that are anyway available over-the-air, without requiring additional sensors, protocols or signals. In principle, this can be achieved by resorting to RSSI-based methods from received *power* measurements associated to the communication signals, but this approach is known to be very inaccurate in practice and would unlikely provide any accuracy gain over plain GNSS technology in real scenarios. In this contribution we claim that cooperative (and opportunistic) localization can be implemented from *timing* measurements, by exploiting simply the receiver timestamps recorded by (legacy) receivers on the communication signals and packets transmitted from other mobile and/or fixed nodes. In our scheme, a set of timestamps from multiple nodes are transformed into a set of pseudo-ranges, and the latter are fused with initial positions to improve and extend the localization function without any need of additional equipment, protocol or signal for ranging. The overall dataflow is sketched in Fig. 1.

## 3. System model and methodology framework

### Reference Scenario

We envisage a cooperative and opportunistic localization system where a subset of the participating nodes know their position with a certain initial accuracy, not necessarily equal for all nodes, from their on-board GNSS module. Furthermore, each node communicates wireless with other mobile nodes and/or with infrastructure nodes, for other purposes than localization. We show that such communication signals can be exploited opportunistically to collectively improve the localization accuracy of the nodes. To achieve this goal, it is only required that each node is able to measure the reception time of individual packets (in case of packet-oriented protocols) or some commonly agreed reference signals (in case of continuous transmissions, e.g. frame start in DVB-T) according to its local clock, and that it is able to share with the rest of the system the recorded reception timestamps, the associated packet/signal identifiers and its initial position estimate (if available). No external mechanism is required to synchronize the clock of the individual receivers, i.e., we consider *asynchronous nodes*. Also, we do not make any assumption on the transmission time of every packet or reference signal, which we assume uncontrolled and unknown. In this way, we can cope with *asynchronous transmissions*. In fact, most data protocols are indeed completely asynchronous, and in standard COTS devices transmission timestamps (if at all available) are less accurate than reception timestamps. However, it is quite straightforward to extend the method presented here to make use of any additional information that might be available on the transmission times (e.g. in case of periodic transmissions). Also, for the sake of simplicity we consider a basic scenario where all measured data are somehow gathered at a centralized entity (server or cloud) in charge of performing the computation, and we do not consider here any privacy or security aspect. Variations of the proposed method to perform distributed, secure and privacy-preserving computation are interesting directions for progressing the research.

## Time-based localization with asynchronous nodes

In order to illustrate the principle of the proposed scheme, we need first to introduce a minimum of notation. We denote by $t_i[m]$ the absolute transmission time of the generic $m$th packet sent by transmitting node $i$, and by $r_k[m]$ the *absolute* reception time of the same packet by the receiving node $k$. Node $k$ may not be the intended destination for packet $m$ and it is not necessary for $k$ to decode the whole packet. We assume only that node $k$ overhears the packet and decodes the minimum set of control information (e.g. the header) that allow to identify the source node (e.g., MAC source address) and to discriminate packet $m$ from other neighboring packets from the same source.

Both transmission and reception times $t_i[m]$ and $r_k[m]$ are referred to an ideal absolute reference clock and cannot be observed directly. What can be measured is only the *local reception timestamp*, denoted by $s_k[m]$, according to the local clock at receiver $k$. We denote by $d_{ik}$ the true distance between transmitter $i$ and receiver $k$. We consider a static scenario where the position of all nodes can be considered fixed at the timescale of interest, leaving extensions to moving nodes as a prominent direction for further study.

When nodes $i$ and $k$ are in direct Line-of-Sight (LOS), the transmit and reception times are linked by ($c$ denoting the light speed):

$$r_k[m] = t_i[m] + d_{ik}[m]/c. \qquad (1)$$

The reception timestamp is linked to the absolute arrival time by the following relation:

$$s_k[m] = a_k + (1 + b_k) * r_k[m] + h(r_k[m]) + w_k[m] \qquad (2)$$

wherein $a_k$ and $b_k$ denote respectively the clock bias (temporal offset) and clock drift (frequency offset) of receiving clock $k$. The clock drift is due to deviations of the actual clock frequency from its nominal value, and is typically limited to a few tens of ppm for commercial devices. The temporal function $h_k(\tau)$ ($\tau$ denoting a generic reception instant) accounts for any additional slowly-varying error component, possibly non-linear, caused by short-term clock fluctuations. This error component is highly correlated between neighboring packets arriving at the same receiver within a small interval. Finally, the term $w_k[m]$ captures the residual measurement error component associated to each individual reception event (including e.g. clock quantization, timestamp truncation and any other source of measurement noise) and can be modeled as an independent random variable. Note that eq. (2) focuses exclusively on clock error components. We are not considering at this stage additional sources of *spatial* measurement errors, like multipath and Non-Line-of-Sight (NLOS), which affect also synchronous systems. These will be discussed later in Sec. 4.

For the sake of illustration simplicity, consider a generic scenario with multiple nodes in known positions ("anchors") and a single node ("blind") whose position is unknown and needs to be determined. Every node transmits packets that are overheard and timestamped by other nodes. For every packet traveling from node *i* to *k* we can write one equation in the form of (2), resulting in a system of equations with the following observables:
- Distances between pairs of anchor nodes;
- Reception timestamps $s_k[m]$;

and the following types of of unknowns:
- Clock error components: $a_k$, $b_k$ and $h_k(\tau)$;
- Transmission times $t_i[m]$;
- Distances (ranges) between the blind node and each anchor node;

In order to locate the blind node, we need to estimate the latter, i.e., the blind-anchor ranges. It turns out that from the system at hand we cannot obtain direct ranges but only *pseudo-ranges*, i.e., a set of distances defined up to a common unknown bias term. This is not a problem since there are well-known methods for computing a position from a set of pseudo-ranges, for instance the standard Iterative Least Squares (ILS) algorithm used in GPS [Tsu04].

In order to obtain pseudo-ranges from a system of equations of the form (2) we need to deal with the other unnecessary unknowns listed above. Generally speaking, each unnecessary unknown can be either *eliminated* upfront, by combining (differentiating) selected equations, or it can be *estimated* from the data at hand. In the latter case, it can be either estimated *jointly* with the variables of interest (i.e. treated as a nuisance parameter) or it can be estimated separately and then compensated. Therefore, different resolution approaches are possible depending on how each unknown is treated. Until now, only a few pioneering work have started to formulate and investigate some of these methods. One of the intended goals of the present contribution is indeed to attract further research to explore other possible solutions and estimation methods for the problem at hand.

Some previous work have considered a simplified clock error model by neglecting the clock drift $b_k$ and the slowly-varying component $h_k(\tau)$. In this case eq. (2) simplifies as shown in Fig. 2-A. One possible resolution approach is to differentiate the equations relative to two packets from different transmitters arriving at the same receiver in order to get rid of the clock bias $a_k$ (ref. Fig 2-B) then treat the transmission times as nuisance parameters. If the two packets are close in time, such a differentiation would cancel also the slowly-varying component $h_k(\tau)$, if present. Alternatively, one can differentiate the equations relative to reception of the same packet (from a single transmitter) arriving at two different receivers in order to get rid of the transmission times $t_i[m]$ (ref. Fig 2-C), then treat the clock bias as nuisance parameter. These two approaches have been considered recently in [Col14] and [Ban14] respectively. However, the simplest approach is to apply a *double differentiation* on the four equations between a pair of transmitters and a pair of receivers, as shown in Fig. 2-D, in order to get rid of both types of unknowns. This method has been termed *Differential* Time-Difference of Arrival (DTDoA) and should be distinguished from the classical Time-Difference of Arrival (TDoA) adopted in GPS. The requirement that anchor nodes (e.g., satellites) be synchronized applies to TDoA but not to DTDoA.

The DTDoA equation appeared a decade ago in the field of interferometric localization [Mar05] and was later considered in systems where time-differences are measured by means of waveform correlators [Fan07]. The idea to use DTDoA with simple timestamp measurements from legacy wireless devices was presented and tested only recently in [Nag11] and [Fac14], where different approaches were adopted to estimate and compensate for the clock drift, among other methodological differences. Interestingly, both work report experimental validation results with COTS IEEE 802.11b devices achieving sub-meter accuracy in LOS conditions. Further work is needed to explore more advanced estimation techniques, however the early result in [Nag11] and [Fac14] indicate that sub-meter accuracy is at reach nowadays with *current* COTS equipment, at least in LOS conditions.

**System model**

From a system-level point of view however, the closest work is represented by [Col14]. The system model considered therein is a particular case of the more general scenario envisioned here. In general, we consider three types of nodes:
- **Fixed stations**: terrestrial infrastructure nodes, whose positions are precisely known, i.e., with negligible positioning error.
- **Mobile nodes with GNSS**, whose positions are known only approximately, i.e., with an initial error of several meters that we aim at reducing.
- **Mobile node without GNSS** (blind node), whose position is completely unknown and needs to be determined.

Depending on the radio technology, fixed stations can be transmit-only (e.g. DVB-T broadcasting towers or LTE beacons), receive-only (e.g. passive WiFi sensors) or both transmitters and receivers, e.g. Road-Side Units (RSU) transceivers. The mobile nodes (vehicles) are assumed to be equipped with terrestrial communication transceivers and able to transmit to and receive from other nodes, both fixed (V2I links) and/or mobile (V2V).

In [Col14] the authors have considered a particularized scenario where fixed stations are transmit-only and the single blind node is receive-only. The other mobile nodes, called "helpers" therein, know their position with certain limited accuracy, and these (inaccurate) positions are given in input to the localization algorithm along with the reception timestamps. In other words, we have two sources of uncertainty (or noise): in the timing measurement and in the helper position. Remarkably, the simulation results presented in [Col14] show that, under certain conditions, *the final error on the blind node position is smaller than the initial error on the helper node positions*. In other words, the output position (of the blind node) is more accurate than the input positions (of the helper). The "accuracy gain" is obviously due to the additional information contributed by the terrestrial timing measurements. This result shows that, in principle, one can further refine the initial positions of the helper nodes (obtained by GNSS) *by exploiting the terrestrial pseudo-ranges information embedded in the reception timestamp data* – instead of additional direct ranging measurements as envisioned in [Gar12]. As noted there, the positioning refinement process can be performed *sequentially*, by feeding back the new (more accurate) position into the estimation problem in order to refine the position of another node and so on, or *jointly*, via more compact estimation procedures that seek to resolve all node positions jointly at once.

## 4. Next steps

The preliminary simulation result presented in [Col14], obtained in a simplified scenario, should be taken as a proof-of-concept indicating a direction for further experimentation rather than a conclusive point. We do not claim at this stage that the exploitation of terrestrial timestamp data by means of asynchronous localization algorithms will *always* improve over initial GNSS accuracy, but we claim that it has the potential of doing so *under certain conditions*. The research question for the academic community is therefore:
(1) Under which conditions terrestrial timestamp measurements complementing GNSS can improve positioning accuracy down to sub-meter level?

And for the industry:
(2) What can be done, e.g. on the side of radio receiver design, to facilitate achieving those conditions?

The goal of this contribution is to solicit academy and industry to pick these research questions as stimuli for progressing further with theoretical analysis and field experimentation. Hereafter we indicate some desirable action points for both communities.

### Commoditization of accurate timestamping.

The attractiveness of the proposed method for practical applications ultimately depends on the quantitative impact of measurement errors into the final position estimation. Generally speaking, we can identify the following sources of errors:
- ***Clock errors***, due e.g. to the (in)stability of clock frequencies and timestamp quantization, among other factors.
- **Multipath errors** in Line-of-Sight (LOS) conditions.
- **Non-Line-of-Sight (NLOS) errors**.

Recall that only clock errors were considered in eq. (2), since the focus there was to show that node synchronization can be effectively replaced by timestamp data processing. The other two error components, namely multipath and NLOS, can be modeled as further additive terms to eq. (2). However, it should be noted that while the clock error components vary in *time*, the multipath and NLOS components vary in *space*.

In order to exploit the potential of cooperative opportunistic localization, as augmentation of or extension to GNSS systems, the collective effect of all these error sources must be kept low.

We have shown that, in principle, clock errors can be counteracted by data processing in the initial estimation phase (ref. Fig. 1). This is achieved by considering multiple measurements in temporal diversity, that is, averaging over multiple packets. However, the adoption of more stable receiver clock and more accurate timestamping capabilities would reduce the power of measurement noise in input, thus facilitating the task of the processing stage and allowing to use a lower number data points (i.e., packets), with beneficial effect also on the estimation delay and computational complexity. When a GNSS module is co-located with a terrestrial wireless receiver (e.g. in OBU or smartphone), the GNSS signal can be passed to the latter in order to control the clock signal used by the timestamping function, thus improving the timestamp accuracy.

Similar considerations apply for the multipath error that is encountered in LOS links due to the composition of multiple signal replicas arriving after the first direct path. The magnitude of this error can be up to a few meters for each individual ranging measurement (see e.g. [Exe12]). It can be counteracted in the estimation phase by leveraging *spatial* diversity and redundancy, i.e., considering an increased number of anchors. If the blind node is in motion, *spatial* diversity is transported into *temporal* diversity, and the same goal can be achieved by considering multiple packets from the same (smaller) set of anchors. Also in this case it would be beneficial to reduce *ex ante* the multipath error of each measurement data point by adopting specific signal processing techniques in the receiver, e.g. to discriminate and timestamp the *first* arriving signal replica, while still leaving the rest of the receiver chain tuning on the *strongest* replica for packet decoding. These algorithms have been well studied in the realm of GPS technology and could be reused in terrestrial wireless communication devices, probably with a relatively minor *marginal* cost in terms of implementation and resource consumption.

We remark however that the experimental results reported by [Nag11] and [Fac14], obtained in LOS conditions with COTS WiFi devices that do not implement any mitigation strategy against multipath and/or clock errors, indicate that even without such additional improvements sub-meter accuracy is at reach nowadays with standard devices.

### Robust estimation algorithms for NLOS conditions

The most serious source of error in our scenario is undoubtedly NLOS reception. In ranging measurements NLOS paths introduce (possibly large) positive bias. NLOS mitigation strategies are an active field of research and already cover a wide range of approaches (see e.g. [Guv09], [Vag13] and references therein). However, most previous work has addressed NLOS problem in the context of *ranging* techniques, while our framework is based on *pseudo-ranges* (or equivalently, range differences). Developing robust estimation methods that can cope with NLOS scenarios is definitely one of the most compelling research challenges in the progress of this work. However, in certain C-ITS applications one can expect that a sufficient number of other nodes are anyway in LOS conditions, e.g. neighboring vehicles and RSUs.

One could think that the "commoditization" of more accurate receiver timestamping, i.e., the implementation in COTS devices of clock error and multipath error mitigation techniques, would be irrelevant considering that NLOS errors, which cannot be resolved by these techniques, are the dominating error source in practical settings. In other words, why attacking the "small" errors (clock and multipath) while the "large" ones (NLOS) are there?

We argue instead that the "commoditization" of accurate receiver timestamping would be beneficial also towards the goal of reducing NLOS errors. First, we should distinguish between "heavy NLOS" and "moderate NLOS" conditions, depending on whether NLOS links are the majority or not. In moderate NLOS scenarios it is not difficult to identify and exclude the few NLOS links with pure data-driven approaches. Furthermore, the reduction of the measurement noise for LOS links will intrinsically facilitate the job of NLOS mitigation techniques, as the subset of LOS measurement will yield a higher degree of internal consistency, making it easier to tell apart the "outlier" NLOS data. Third, improving the achievable localization accuracy in favorable conditions (LOS or moderate NLOS) down to decimeter level, would certainly motivate additional experimentation and pilot deployment of communication system with built-in localization functions. This in turn bears the potential to attract more relevant research work on robust estimation methods in adverse NLOS conditions, triggering a positive virtuous cycle.

### Extension to tracking

For the sake of simplicity we have presented the fundamental principles of the proposed method with reference to a static scenario. Extension of the proposed model to dynamic scenarios, where some of the nodes are in motion, is an important direction for further research. Going from static localization to dynamic tracking implies certainly an increase of computational complexity, as pseudo-ranges and the other parameters of interest vary in time. However, since node mobility increases both spatial and temporal diversity, it should be advantageous in terms of achievable accuracy. This is true for pure GNSS-based localization, and a fortiori for hybrid GNSS/terrestrial localization.

## 5. Conclusions

We envision an increasing level of integration between GNSS positioning and terrestrial radio localization, with the latter augmenting and complementing the former. Given the ubiquitous presence of legacy radio signals and the widespread of multi-radio devices, it is appealing to devise methods to reuse available signals and devices for localization purposes. We have shown that this can be achieved in principle by relying simply on receiver timestamps, with no need of additional ranging methods or protocols. A few pioneering work have demonstrated that time-based localization in asynchronous networks can achieve sub-meter accuracy. Waiving the requirement of node synchronization is an important step towards unlocking the potential for opportunistic and cooperative radio localization in combination to GNSS, which we hope will motivate further research and experimentation in this direction.

## Acknowledgments

The author would like to thank (in alphabetical order) Angelo Coluccia, Nicoló Facchi, Francesco Gringoli, Giuseppe Ricci and Andrea Toma for several fruitful discussions about resolution approaches for time-based localization in asynchronous networks.

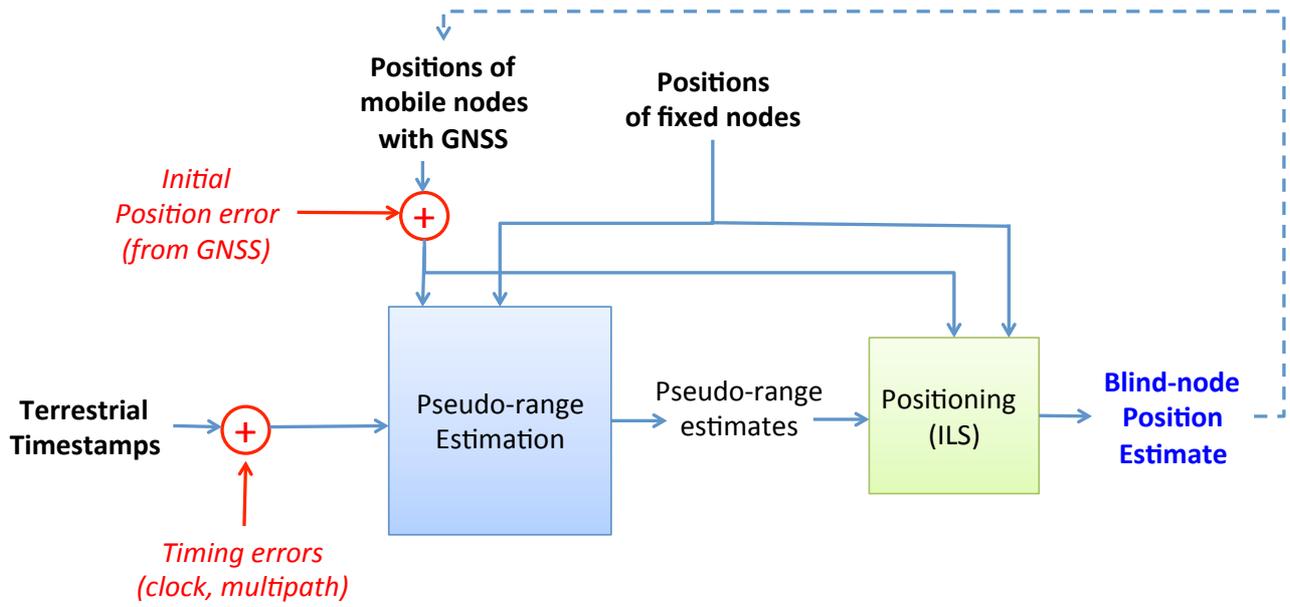

**Figure 1** – Conceptual workflow diagram of a two-stage localization procedure based on reception timestamps.

| | Transmitters | Receivers | Equation (simplified) | References |
|---|---|---|---|---|
| A. | $i$ —$m$→ | $k$ | $s_k[m] = a_k + t_i[m] + \dfrac{d_{ik}}{c} + noise$ | |
| B. | $i$ —$m$→, $j$ —$m+1$→ | $k$ | $s_k[m+1] - s_k[m] =$ $= t_j[m+1] - t_i[m] + \dfrac{d_{jk} - d_{ik}}{c} + noise$ | [Col14] |
| C. | $i$ —$m$→ | $k$, $l$ | $s_k[m] - s_l[m] =$ $= a_k - a_l + \dfrac{d_{ik} - d_{il}}{c} + noise$ | [Ban14] |
| D. | $i$ —$m$→, $j$ —$m+1$→ | $k$, $l$ | **DTDoA** $s_k[m+1] - s_k[m] + s_l[m+1] - s_l[m] =$ $= \dfrac{d_{jk} - d_{ik} + d_{jl} - d_{il}}{c} + noise$ | [Nag11] [Fac14] |

**Figure 2** - Compact overview of different approaches for asynchronous time-based localization considered in recent literature. The DTDOA procedures presented in [Nag11] and [Fac14] include also clock drift compensation (not shown in the equation).